# Energy systems models are diagnostic tools, not projection machines

I. David Elder[1,7,*], Juan Moreno-Cruz[2], Cameron Wade[3], Sylvia Sleep[4], Sara Hastings-Simon[5], Sean McCoy[6], Heather L. MacLean[1], I. Daniel Posen[1,8,**]

[1]Department of Civil & Mineral Engineering, University of Toronto, Toronto, Canada
[2]School of Environment, Enterprise and Development, University of Waterloo, Waterloo, Canada
[3]Sutubra Research, Halifax, Canada
[4]Sustainable Systems Engineering, University of Calgary, Calgary, Canada
[5]Department of Earth, Energy, and Environment, University of Calgary, Calgary, Canada
[6]Department of Chemical and Petroleum Engineering, University of Calgary, Calgary, Canada
[7]Lead contact
[8]Senior author
*Correspondence: ian.elder@mail.utoronto.ca
**Correspondence: daniel.posen@utoronto.ca

## Abstract

Energy systems models (ESMs) are a leading tool to guide the energy transition. They have been influential in supporting national decarbonisation strategies and regional system planning, but these complex models can be opaque. Their results are often presented predictively or prescriptively, with little exploration of uncertainty and without clear discussion of limitations. Consequently, projections of ESMs have often been given more authority than their evidence can bear, undermining their contributions to energy transition policy.

We argue that ESMs should instead be applied as explanatory, diagnostic tools. Model studies should explore uncertainty to find robust insights and define limitations, interrogate model behaviour to find testable real-world explanations, and communicate these explanations plainly and responsibly. This approach shifts the evidentiary burden from plausible projections to real-world insights that can be broadly understood and debated, and considered alongside other forms of evidence. Used this way, ESMs can support robust, justifiable, and pluralistic decision-making.

# From projections to justified insights and pluralistic decision-making

Quantitative models and their results are often given a privileged position in decision processes[1]. Among these, energy systems models (ESMs) have emerged as a leading decision-making tool for the energy transition[2,3]. These complex computer models are valued for their capability to model and inform infrastructure investment, system operation, policy design, and long-term planning under technical, economic, and environmental constraints and uncertainty. The world invests more than $3 trillion USD each year in the global energy system[4], which is expected to increase as we transition to net zero emissions[5]. When models inform decisions at this scale, we should carefully consider how insights are justified. We argue that too great an evidentiary burden has been placed on the projections of ESMs, particularly in the case of long-term planning models. These projections have, in many ways, anchored the way we think about the energy transition[6,7], yet ex-post analyses suggest that, due to unforeseen outcomes, reality has diverged significantly from these projections[8–10]. We need methods of justification that relieve the evidentiary burden from uncertain projections of the future.

Overconfidence in ESM projections can distort decision-making in two directions. Where projections are overly optimistic, they may encourage decision-makers to underestimate future barriers, inequities, or alternative pathways, contributing to infrastructure lock-in, policy backlash, stranded investment, or delayed mitigation. Biofuels, hydrogen, and carbon capture and storage illustrate this risk. These technologies have appeared prominently in modelled decarbonisation pathways[11], but their real-world deployment depends on factors that are often only partially represented in ESMs[12]: infrastructure expansion, land-use constraints, public acceptance, and behavioural decision-making. If study limitations are not communicated clearly, projections can be mistaken for evidence of feasibility. Conversely, where projections are overly pessimistic, they can discourage investment in technologies that later prove important. Cost and performance improvements in solar photovoltaics and batteries have repeatedly outpaced expectations. Once treated as peripheral[13,14], these technologies have become central to decarbonisation planning. We do not suggest that these outcomes were foreseeable; the issue is not that models fail to predict the future, but that their projections are often granted authority as if they could[15]. Awareness of these issues can also inspire the opposite reaction, where decision-makers lose trust and dismiss the models entirely[16].

Under deep uncertainty, the insights of ESMs should derive authority not from plausible projections but from diverse and rigorous testing of systems intuitions, applying projections as a mode of operation rather than the final insight. The energy transition is an impenetrably complex problem, rife with uncertainties, where we lack clear common objectives—a *wicked problem*, which cannot be "solved" from a single, limited perspective[17]. Wicked problems require *pluralistic* decision-making, where we encourage the participation of diverse perspectives and expertise to test insights more broadly[18,19] and to justly and democratically balance competing objectives[17,20]. This process requires communicable, falsifiable insights that can be understood and tested from as many perspectives as possible (spanning stakeholders, experts, and research methods), and which are presented to decision-makers with a clear definition of their *epistemic limits*—the limits of what can be justifiably known based on available evidence (to elucidate the meaning of this term in relation to energy systems modelling, we present an analogy in Section S1).

To guide ESMs towards justifiable insights in support of pluralistic decision-making, we propose a framework of principles for modelling practice called Diagnostic Modelling, building on a broad literature of exploratory and open modelling, robust decision-making, modelling for policy, and theories of scientific inference. When modelling under this framework we perform *diagnostic* analysis as opposed to *predictive* or *prescriptive* analysis (Table 1), shifting the evidentiary burden of modelling from projections to testable explanations.

***Table 1. Four types of analysis in the context of ESMs.*** *Hypothetical example insights. Adapted from Sarker (2021)[21].*

| Type | Question | Insight |
|---|---|---|
| Descriptive | "What happened?" | "When we applied a policy offsetting investment risk, the model built 20-50% more onshore wind." |
| Predictive | "What will happen?" | "Policies offsetting investment risk would increase onshore wind capacity by 20-50%." |
| Prescriptive | "What should we do?" | "Governments should implement policies offsetting the risk of onshore wind investment." |
| Diagnostic | "How does it work?" | "Policies offsetting risk reduce interest rates and make investment in onshore wind more cost-effective." |

We contend that, in the case of energy systems modelling, descriptive insights only describe model behaviour, not real-world behaviour, that predictive insights are inherently unreliable, and that a model cannot justifiably prescribe what society should do. The strength of an ESM is therefore diagnostic: to explore, interrogate, and communicate mechanisms that build an intuitive understanding of the decision space and its trade-offs. In Diagnostic Modelling, three simple principles guide this process and help frame discussions around the validity of insights (Figure 1):

**Exploration**. Studies should responsibly consider uncertainties that might affect their results or insights. They should explore these uncertainties to define the limitations of the study and, within these limits, robust insights that inform decisions for an uncertain future.

**Interrogation.** Studies should interrogate their model, testing intuitions and seeking mechanistic explanations for model behaviour that directly translate into real-world insights. Authority is derived not from the quality of the model, assumptions, or projection but from the strength of the explanation and the severity and diversity of its testing.

**Communication.** Studies should primarily communicate real-world insights rather than descriptions of model results. The mechanisms underlying these insights should be explained clearly and plainly for a broad audience, responsibly bounded within epistemic limits, leaving remaining unknowns to the mandate of pluralistic decision processes.

# EXPLORATION

Explore all types of uncertainty to define limitations and find robust insights

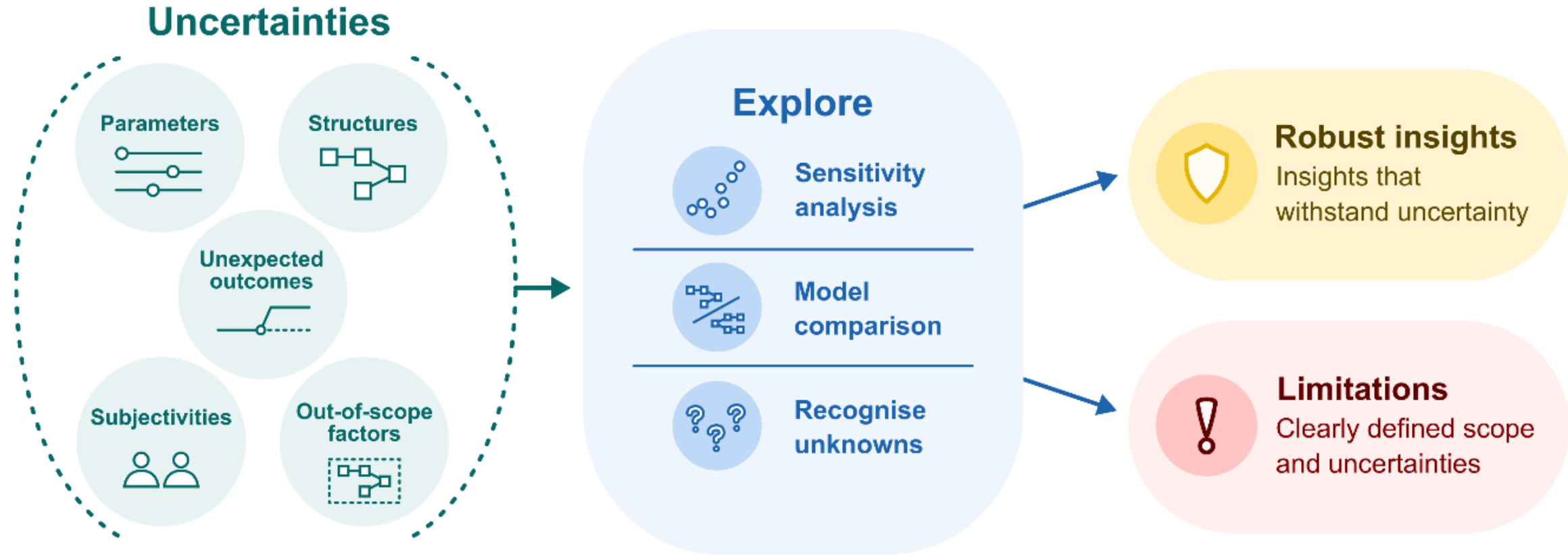

# INTERROGATION

Construct a real-world explanation for your result then test it

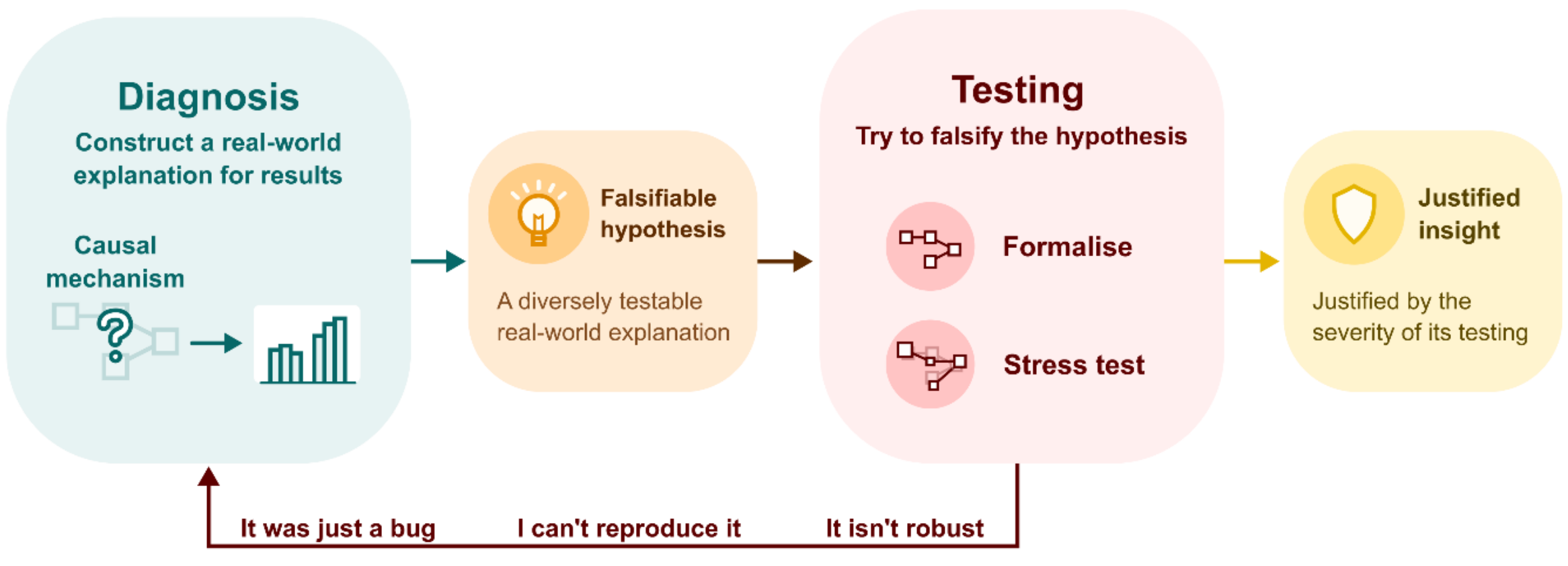

# COMMUNICATION

Clearly communicate limitations and falsifiable explanations

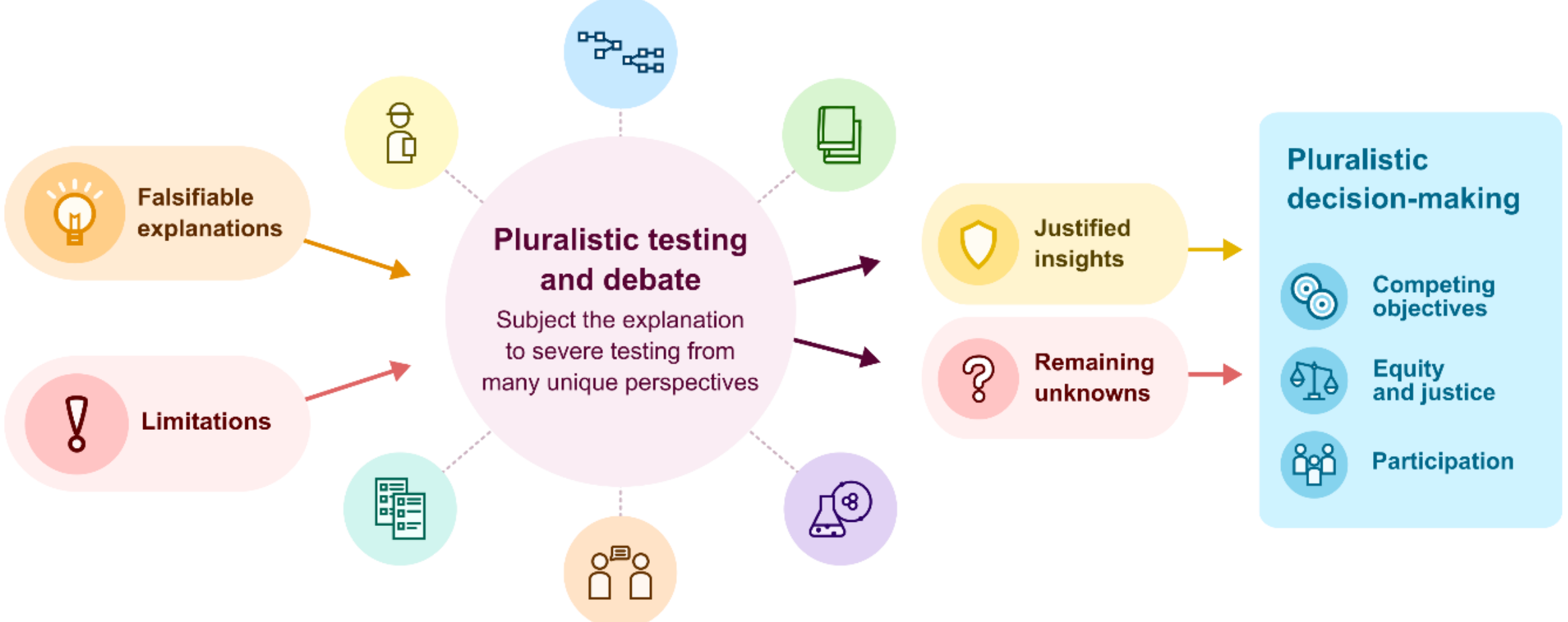

***Figure 1. The three principles of Diagnostic Modelling.*** *Exploration defines limitations and finds robust insights that withstand uncertainty. Interrogation seeks mechanistic explanations for model results so that these explanations can be diversely tested and justified. Communication presents explanations and limitations to the broader forum for further testing and debate, supporting pluralistic decision processes.*

Elements of Diagnostic Modelling may be familiar: many ESM studies already explore uncertainty, interrogate model behaviour, or communicate limitations, as we demonstrate with exemplars in Section S2. Yet we are not aware of a single study, including our own, that brings all three together holistically. Diagnostic Modelling treats this not as a personal failure of modellers, but more broadly as a sign that ESMs still lack well-defined, tractable, and scientifically defensible methods of justification to guide modelling practice.

# Explore uncertainties for robust insights and to define limitations

Energy systems models have become increasingly complex in the pursuit of model realism, but complex models are less interpretable, less explorable, and more uncertain[22,23]. To find robust insights for an uncertain future, exploration of uncertainty should be prioritised over model completeness or complexity. We argue, however, that it is impossible to model all the uncertainty of the energy transition. In Diagnostic Modelling, the primary objective of exploration, therefore, is not to model all uncertainty but rather to clearly define the limitations of the study and its insights.

Despite an oft-touted goal of insights-not-numbers[24–26], ESM studies have historically employed a prescriptive or predictive, scenario-based approach[27,28] with limited exploration of uncertainty[28]. However, this has improved with advances in uncertainty methods and practices[28] (see Table S2 for a brief review of existing uncertainty methods for ESMs). ESMs contain many different types of uncertainty but most studies that do explore uncertainty model only parametric uncertainty (uncertain input data)[28]. Increasing recognition has also been given to structural uncertainties[28] (uncertain choices in the structuring of the model), though methods to address structural uncertainty are more nascent. These are further complicated by subjectivities and biases in the construction of the model or study and in its interpretation by the modeller and the audience (including, for example, anchoring bias, logical fallacies, misinterpretation, or disagreement on objectives[29]). We explore a taxonomy of ESM uncertainties in Section S3 to identify underexplored uncertainty types (e.g., epistemic uncertainties, subjectivities), though different types of uncertainties are, in practice, often inseparable in modelling.

A major challenge in energy systems modelling is the high dimensionality of uncertainty, which makes complete exploration impossible. To illustrate, consider N independent uncertain variables taking one of three values each, with equal probability (e.g., low, central, or high). As N increases (as we increase the dimensions of the scenario space), the number of possible scenarios grows exponentially and, counterintuitively, possible scenarios are found further from the centre on average—i.e., the bounds of the scenario space grow outward and the region of the scenario space near the central scenario "empties out" (Figure 2A). In other words, as we model more uncertain variables, our central projection becomes less representative of the scenario space (Figure 2B). We expand on the geometric explanation for this phenomenon in Section S4 but note that this thought experiment is highly simplified. In practice, many uncertainties may be bounded or correlated, reducing effective dimensionality, or the model may be insensitive to them. For a more-sophisticated analysis of how complexity increases model uncertainty, see Puy et al. (2022)[22].

If we sample each variable from a normal distribution, instead of our three discrete values, we should also expect an increasing occurrence of extreme tail outcomes (Figure 2C). These outcomes are not peripheral. A single unexpected shift can restructure the decision space, as solar photovoltaics illustrate: largely absent from ESM projections at the turn of the century[14], they are now central to system planning. Looking forward, model insights should be judged not only by their plausibility under central assumptions, but also by their robustness to a range of similarly disruptive developments.

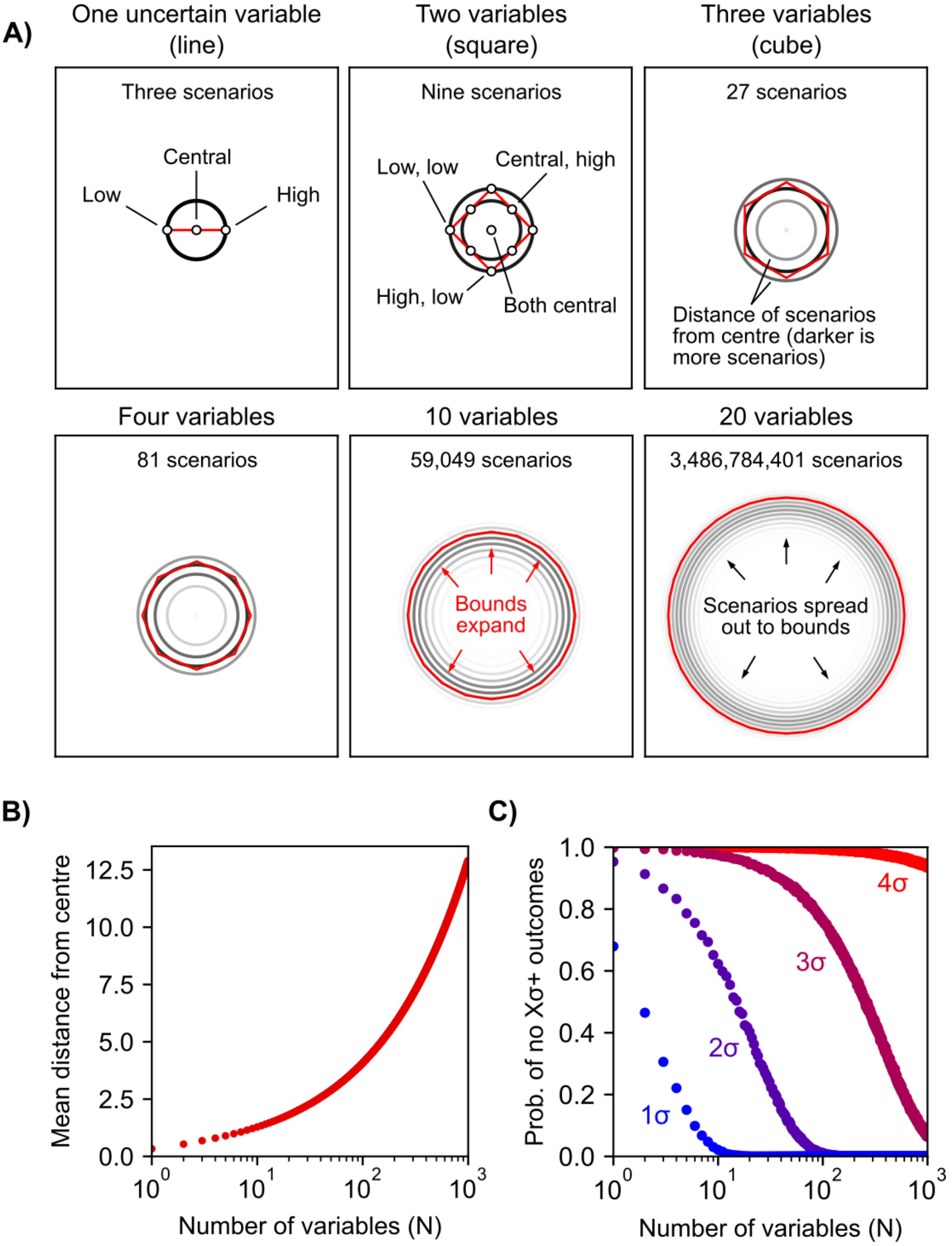

***Figure 2. Increasing uncertainty with N independent uncertain variables.*** *Generated by Monte Carlo sampling of N variables each sampled from {0, 0.5, 1} with equal probability. (A) In higher dimensions, the number of possible scenarios grows exponentially as* $3^N$ *and most scenarios are found farther from the centre. Red outlines are Petrie polygons: 2D projections of each scenario space (whose full, N-dimensional geometry is a hypercube) roughly showing the outer bounds of the space. Gray rings show the frequency of scenarios found at different radial distances from the centre. (B) The more uncertainties in our model, the less well the central scenario represents all possible*

*scenarios. Red points show mean distance of all possible scenarios from the central scenario as the number of uncertain variables increases. C) The more variables in our model, the more likely it is that at least one will be realised with an extreme outcome. Probability that none of N independent uncertain variables, each sampled from a normal distribution, instead, are realised beyond 1σ (one standard deviation, 32% probability), 2σ (5%), 3σ (0.003%), or 4σ (1 in 10,000). With N=250 independent uncertain variables (a relatively small number in complex systems), the probability of at least one 3σ outcome is about 50/50. For 2σ, this occurs at N=15. Note semi-log axes. See Section S4 for additional explanation.*

Complexity increases not only scenario uncertainty but also uncertainty of interpretation, making insights more difficult to discover or communicate, and of model operation (data entry mistakes, code bugs) while increasing computational cost, limiting the feasible scope of exploration. This "curse of dimensionality" contradicts a common attitude in modelling where modellers must improve their forecast accuracies (the "goodness" of their model) by increasing model complexity[28,30,31]—adding more independent uncertain variables at the expense of exploration and interpretability[6,28]. This is often framed as a necessity to faithfully represent the system[32,33]—a reduction of *completeness uncertainty* which is given undue priority over other uncertainty types[28,33].

Certainly, there is a need for completeness—or even complexity[34,35]. In fact, it is the complexity of models that often produces novel insight, found in the emergent phenomena of many interacting assumptions[36]. Increased complexity can result in more-sophisticated—and so, we hope, more surprising and insightful—phenomena. However, these phenomena may be increasingly difficult to interpret. Complexity can be a source of insight but we must dispel the idea that it is a treatment for *uncertainty* when, in fact, it is a root cause[23,37]. For each model there exists some crossover point where the benefits of increasing complexity are overwhelmed by lost exploration (due to limits in compute resources, time, and effort), decreased interpretability and compounding uncertainties[24,38]. So, we need complexity for insightfulness and realism, but we also need simplicity, meaning smaller, more tractable, more interpretable models, used to explore more of the scenario space and find robust insights that withstand uncertainty[37,39,40].

However, system uncertainties exist whether we model them or not. Our number of uncertain variables (N) is not a function of the number of variables in our model but rather the number of independent uncertainties in the system we are modelling. From this follow three insights: (1) to computationally explore all uncertainty, we need not just a simple model but a model of a simple system, such that all intrinsic uncertainties can be fully defined and modelled (if such a thing is even possible); (2) aggregating variables may enable a greater exploration of uncertainty (due to reduced computational intensity) without changing the intrinsic uncertainties of the study; (3) for models of large, complex systems—where the scope of the model cannot capture all the intrinsic uncertainties of the system—an uncertainty modelling exercise (e.g., parametric sensitivity analysis), though necessary, may be insufficient.

To demonstrate the benefits of a breadth-first approach to modelling uncertainty, we point to Sepulveda et al. (2018)[41] and Neumann and Brown (2023)[42], whose efforts we discuss in more detail in Section S2.1. However, following our third insight, we stress that it is not possible or appropriate to model all types of uncertainty. Modelling several hundred scenarios can generate valuable and robust insights, as these studies have shown, but the required effort may be impractical and the apparent exhaustiveness may give the sense that uncertainty has been fully explored, which is not possible. The most influential uncertainties may not be parametrizable or within the structural scope of the model, requiring qualitative discussion or, where implications are also deeply uncertain, simply acknowledgement so that conclusions are not given undue authority[43]. Mersch, Markides, and Mac Dowell (2023)[44], which we also discuss further in Section S2.1, demonstrate this latter approach, expounding on many unmodelled uncertain factors that limit their insights. In Diagnostic Modelling, uncertainties are explored not as a sensitivity exercise but rather to find robust insights for an uncertain future and, perhaps more importantly, to define their epistemic limits—the boundary between results and justifiable insights, between the model world and reality.

## Interrogate models for testable real-world explanations

Where uncertainty cannot be fully modelled or explored, we require a new paradigm of justification. In Diagnostic Modelling, we interrogate the model for *diagnostic* insights (see Table 1). These are mechanistic, cause-and-effect explanations (or *causal explanations*) of model behaviour, in real-world terms (not in terms of model structures), that withstand uncertainty and remain robust to model assumptions. These mechanisms can be explained plainly, to a broad audience, justified solely by the reason of that explanation and the rigour of its testing, without the need to explain or justify the model itself. Explanations are communicable, falsifiable, and less dependent on their originating model, making them more robust to uncertainty.

A model is not the real world and model experiments do not produce real-world observations[45]. In experiments such as we might see in physics, chemistry, medicine, or a well-designed survey, the results themselves contain a real-world observation prior to any analysis or interpretation. Not so in modelling, where results could be a bug or an artefact of model assumptions, structure, or scope. Seeking explanations pushes us to think again in real-world terms.

Real-world explanations also support a pluralistic approach to decision making, as they are easier to communicate to a broad audience and can be tested by other models or by those outside the field of energy systems modelling, leaving room in the decision-making process for stakeholder preferences and consensus from other studies and disciplines[40,46,47]. They may also be more generalisable and enduring. Mechanisms describe general system structures rather than outcomes for a particular instance of a system in one time and place. They may therefore be generalisable to other times and places, or other systems. By accumulating these insights, we incrementally refine the collective mental models we use each day to interpret the messy world around us and make executive decisions under complexity[43]. This approach benefits from a key strength of ESMs: their bottom-up structure, grounded in formalised system mechanisms, makes them structurally diagnosable.

To justify diagnostic insights, the modeller must reject easy conclusions and ask, "Why?" Rather than accept model outputs, interrogate them until insights emerge as explainable real-world mechanisms. Form a hypothesis (a possible interpretation of results) and seek, in earnest, to disprove it. ESMs suffer from two problems that make formal statistical inference unreliable. First, their uncertainties, and therefore probabilities, cannot be fully quantified or explored (as we have already argued). Second, it is impossible to compare their projections to future observations because those observations have not yet occurred[27]. As a result, energy systems modelling often relies on a heuristic form of inference, based on how plausible the hypothesis appears, the apparent falsifiability of the hypothesis, and how much rigour has been applied in attempting to falsify it[18,19,48]. Through interrogation, Diagnostic Modelling seeks to justify hypotheses by maximising their falsifiability and the diversity and severity of their testing.

Interrogation has two modes: diagnosis and testing (Figure 3). In *diagnosis*, we seek to explain the underlying behaviour of results, leading to novel and surprising insights. These emerge from unexpected intermediate interactions between model assumptions (emergent phenomena) that can be translated into mechanistic understanding of the real-world energy system.

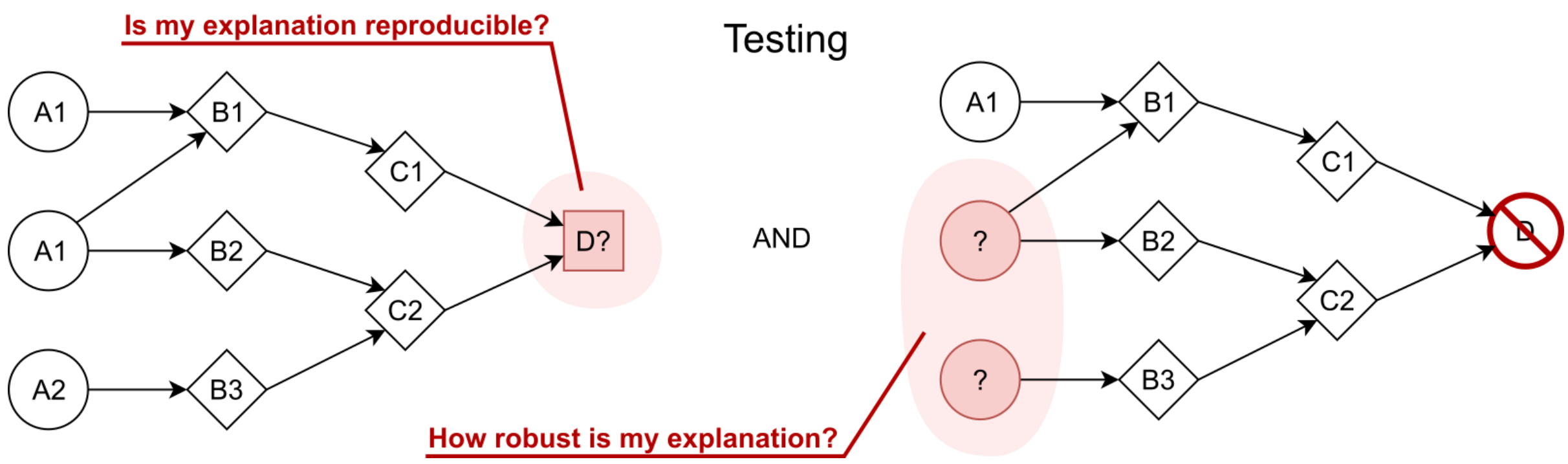

***Figure 3. Diagnosis and testing modes of interrogation.*** *In diagnosis (top figure), we seek unexpected emergent phenomena that, if explained, might translate to real-world insights. Input assumptions A1-A3 produce a model outcome D. Surprising insight emerges because the causal chain leading from assumptions to outcomes (intermediate interactions, question marks) is not obvious beforehand. Once the analyst can diagnose which model structures are responsible for the outcomes of interest, an intuitive, real-world explanation may emerge—a causal description* ($A \rightarrow D$) *becomes a causal explanation* ($A \Rightarrow B \Rightarrow C \Rightarrow D$)*. Note that additional intermediate interactions between those shown here may be sought ad infinitum. Diagnosis ends when the intermediate interactions are explainable not as abstract model structures but as a complete chain of real-world mechanisms. In testing (bottom figure), we test our proposed explanation, seeing if we can formalise and robustly reproduce it (left) and under what assumptions it holds (right).*

As an example of the interrogation process, consider Cano Renteria, Schwartz, and Jenkins (2025)[49]. In Section 3.2.1, they interrogate the hourly operational decisions of their model to construct a robust mechanistic explanation for value synergies between advanced nuclear fission reactors and thermal storage. They identify a first-order intermediate interaction—that fission plants with thermal storage displace more wind, batteries, and natural gas plants, but this is still a description of model outputs, not a real-world explanation, so they interrogate further, seeking a second-order interaction. By considering the operational behaviour of the nuclear plant, they construct a detailed mechanistic explanation in real-world terms: in short, the nuclear plant can oversize its turbine and use thermal storage to carry baseload output power to high load hours, leaving more net load in the middle of the day to be picked up by cheap solar PV and offsetting the need for expensive lithium batteries to perform this arbitrage. This hypothesis can be explained alone, as we have just done, without the burden of explaining or justifying the rest of the study or the model. It can therefore also be reproduced or falsified alone, without the need to discredit the underlying study and model[48].

In *testing*, which may test either prior intuitions or insights emerging from diagnosis, we seek to formalise and falsify explanations. Test each of these statements, especially when tempted not to:

1. "This is a result of a bug in my model. If I fix that bug, the outcome changes."
2. "This is a result of my own input assumptions or data. Different, reasonable assumptions do not reproduce this outcome."
3. "This is a result of how I chose to represent these systems. I can represent these systems in a reasonable alternative way such that a different outcome is produced."
4. "This is an artefact of computer modelling or hidden within a black box. I cannot reasonably justify or explain this outcome in terms of a real-world mechanism."
5. "I have not adequately considered alternatives. If I try, I can justify a different explanation."

Testing these statements requires diagnostic analysis of the model, which requires diagnostic tools. These are tools which open wide the black box to analyse and display the underlying mechanisms of the model, so that they can be evaluated for real-world relevance. We believe there is a need to develop and propagate new diagnostic tools for ESMs, and to better apply existing ones. These might be modelling methods or protocols, methods to explore uncertainty, or methods of visualisation to communicate model dynamics. As a starting point, we discuss some existing tools for Diagnostic Modelling and some examples of their application in Section S5.

An insight that survives interrogation should be real-world relevant and explainable without the model that produced it, making it broadly testable. The more testing it withstands, the more justified the explanation becomes[18,48]. Then, rather than present results to decision-makers and attempt to justify the underlying model or projection, we present an explanation that they can understand and test against their own intuition, justified by all the testing that explanation already withstood.

## Communicate explanations and limitations plainly and responsibly

In Diagnostic Modelling, we recognise that justification requires diverse consensus (pluralism). Modellers should not attempt to provide complete answers to wicked problems but instead leverage the strengths of their specific methods, models, or expertise to communicate bounded insights to the wider forum. Explanations, diagnosed and tested through interrogation, should be communicated plainly and responsibly within the epistemic limits of the study, defined through exploration of uncertainties. This approach encourages pluralistic debate, testing insights from many diverse perspectives for stronger justification, and makes clear to decision-makers what is actually known and what remains unknown, aligning the authority of insights to their evidence[16,50,51].

Myriad tools and methods have emerged to assess policies, strategies, and technologies for the energy transition, but each has its strengths and weaknesses. Each one offers a single, limited perspective. No method can capture all the complexity of the energy transition; every method will unavoidably emphasise certain factors over others and neglect numerous dimensions of the problem. Each perspective is limited but valuable—when applied responsibly, thoughtfully, and humbly.

An insight that holds true from one or more perspectives may not hold true from another. The result is that diverse testing with many different methods, both quantitative and qualitative (mixed-methods), is more likely to falsify insights and therefore creates stronger justification[51–53]. It is not the responsibility of each study to, alone, apply all available methods in tandem. Rather, it is our responsibility to be sufficiently aware of alternative methods and their contributions to the forum, so we can contribute to this debate and help synthesise its conclusions. It is the role of the entire field—academic, public, and private—to collaborate, discuss, and democratically debate to test insights and find just decisions for a plurality of interests[20,47]. To demonstrate our own efforts in this regard, we have reviewed some available decision-making methods and tools for the energy transition in Section S6, where, to broaden our own perspective, we attempt to broadly classify some methods and review their claimed strengths and weaknesses in Tables S1-S5.

The challenge of pluralistic decision-making is finding a common language—one that can be understood and debated by different scientific fields, by industry experts, by decision-makers, and by stakeholders. In Diagnostic Modelling, this is achieved through responsible communication of explanations and limitations. To facilitate a pluralistic decision-making process, insights should be (1) presented clearly and plainly so they can be understood and debated by a broad audience, (2) grounded in real world mechanisms—those diagnosed and tested through interrogation—so that they can be justified or falsified by the merit of that explanation alone, without the need to understand the model itself, and, (3) clearly bounded within the limitations of the study—including modelled and unmodelled uncertainties—to guide further testing, to make space for other perspectives, and so that decisions can be made with an informed understanding of what we actually know. To demonstrate, in Section S2.3, we discuss the work of Usher and Strachan (2012)[54] as an exemplar of responsible communication for ESM studies.

## Modelling remains essential when insights are responsibly justified

Energy systems models, if applied responsibly, are likely to remain central to decision-making for the energy transition. They are among few tools capable of representing system-wide interactions of investment decisions, system operations, and policy. However, their value does not lie in accurately projecting decarbonisation pathways but in diagnostic analysis of system behaviour. Modellers should leverage the strengths of ESMs for diagnostic analysis, as opposed to predictive or prescriptive analysis. They should explore uncertainties and interrogate model behaviour to identify robust mechanistic explanations that reflect real-world insights, then communicate these explanations plainly, responsibly bounded by the limitations and uncertainties of the study. We suggest this simple list of final checks for modellers, or for those engaging with ESM studies:

1. Claimed real-world insights are clearly distinguished from descriptions of model results.
2. These insights are explained through real-world causal mechanisms that can be understood and tested without knowledge of the original model or study.
3. Proposed explanations have been subjected to reasonable initial tests that could plausibly have falsified them, though justification may require further testing.
4. Remaining uncertainties have been transparently explored, or at least acknowledged, and the strength of each claim is duly bounded by the resulting limitations of the study.

We recognise that modelling must balance scientific idealism with pragmatism. Decision-making, and decision-makers, often demand straight-forward recommendations on short timelines[55]. So, a strict application of these—admittedly idealistic—principles may not always be pragmatic or possible. These are principles to guide good practice and to frame discussions of justifiable modelling, not a minimum standard to be met in all model studies. Our arguments do not apply equally to all modelling efforts. When constrained by reality, we do what we can, but we should always communicate responsibly and keep clearly in mind what may be lost when we sacrifice our ideals for the sake of pragmatism.

Diagnostic Modelling seeks to ensure the value of energy systems modelling, not diminish it. It shifts the evidentiary burden from uncertain projections to transparent and testable explanations. This aligns the strengths of ESMs with their decision support role: powerful diagnostic tools rather than unreliable projection machines. Insights produced through Diagnostic Modelling are more transparent, justifiable, and responsibly bounded, facilitating open discussion alongside other forms of evidence for democratic and pluralistic decision-making in the energy transition.

# Acknowledgements

We acknowledge funding from: Natural Sciences and Engineering Research Council of Canada Alliance Missions Grants (577-081-2022), the Ontario Graduate Scholarship (I.D.E.), the Lawson Climate & Sustainability Graduate Award (I.D.E.), the C.W. Bowman Graduate Scholarship (I.D.E.), and the Bert Wasmund Graduate Scholarship (I.D.E.). This research was undertaken, in part, thanks to funding from the Canada Research Chairs Program from Government of Canada (CRC-2023-00181, held by J.M-C.; CRC 2020-00131, held by H.L.M.; and CRC-2020-00082, held by I.D.P.). We are grateful for valuable discussion of ideas with T. Rodrigues, C. Fitzgibbon, K. Rankin, F. R. Zetter Salcedo, J. Elder, S. Hossaini, and J.W. Bond.

# Author contributions

Writing – original draft, I.D.E.; Writing – review & editing, and funding acquisition, I.D.E, J.M-C., C.W., S.S., S.H., S.M., H.L.M., I.D.P.; Conceptualisation, I.D.E., J.M-C., H.L.M., and I.D.P.; Supervision, H.L.M and I.D.P. All authors contributed to discussion and refinement of ideas.

# Declaration of interests

The authors declare no competing interests.

# Declaration of usage of generative AI

During the preparation of this work, the authors used large language models (Claude and ChatGPT) to search for relevant research articles, particularly when searching by concepts, and to critique writing. Authors take full responsibility for the content of the publication.

# Supplemental Information

Document S1: Tables S1-S8, Figure S1, and supplemental discussion Sections S1-S6

Supplemental Information:

# Energy systems models are diagnostic tools, not projection machines

I. David Elder[1,7,*], Juan Moreno-Cruz[2], Cameron Wade[3], Sylvia Sleep[4], Sara Hastings-Simon[5], Sean McCoy[6], Heather L. MacLean[1], I. Daniel Posen[1,8,**]
[1]Department of Civil & Mineral Engineering, University of Toronto, Toronto, Canada
[2]School of Environment, Enterprise and Development, University of Waterloo, Waterloo, Canada
[3]Sutubra Research, Halifax, Canada
[4]Sustainable Systems Engineering, University of Calgary, Calgary, Canada
[5]Department of Earth, Energy, and Environment, University of Calgary, Calgary, Canada
[6]Department of Chemical and Petroleum Engineering, University of Calgary, Calgary, Canada
[7]Lead contact
[8]Senior author
*Correspondence: ian.elder@mail.utoronto.ca
**Correspondence: daniel.posen@utoronto.ca

# S1. Epistemic limits of ESMs: the parable of the red buttons

The word epistemic relates to the philosophical field of epistemology, which studies how evidence is used to justify knowledge. In the context of ESMs, epistemic limits are the limits of what we are able to justifiably claim based on the evidence of our modelling efforts, after accounting for associated uncertainties (see Section S2 for a discussion of the types of uncertainties involved in energy systems modelling). Describing the outputs of a model carries little evidential burden but, if we wish to bring a conclusion out of the model world and make claims about the real world, this requires justification. No model can capture the full complexity of reality, which bounds the real-world claims we are able to make.

Consider this thought experiment: you are asked to step into a room and determine the effect of pressing a button. Inside this room, you find a red button waiting to be pressed but, also, far ahead, the number "0" projected on a large screen. When you press the button, the number on the screen increases to "1". You press again and it changes to "2", then again to "3", "4", "5", etc. You reasonably infer that pressing this button adds one to the number on the screen, and turn to leave the room. As you leave the room, however, you notice that yours is not the only room nor the only button. To either side of you are two more rooms with two more red buttons, separated from yours by panes of glass. Beyond those you can see two more rooms with two more red buttons and so on out of sight.

When you leave the room, you are told that the researchers running the experiment will now step into your room and press your red button approximately five to seven times. Then they might step into the other rooms and press those red buttons as well. They ask you, then, when all the button pressing is finished, what can you say about the final number on the screen?

You are confident about the effect of your red button (factors within the scope of your study), but you can only speculate about the effect of the other red buttons (factors you did not study). You also do not know how many other red buttons there are, or how many times they will be pressed (remaining unknown factors). You have limited but justifiable knowledge to contribute to discussion about the number (that your button will probably increase the number by approximately five to seven) but should be wary, given your limited evidence, about predicting the number's final value. In the same way, an ESM study may justify claims about particular mechanisms within its scope without making predictions about the future state of the energy system.

For some further reading on the epistemology of energy modelling, and the epistemic beliefs of energy modellers, we recommend Bock and Pfenninger-Lee (2025)[S1].

# S2. Exemplars for the principles of Diagnostic Modelling

While we could not find a single article that exemplified all three principles, we were able to find several articles to demonstrate a responsible application of at least one principle. We walk through a few of our preferred examples here and explain how they uphold at least one principle of Diagnostic Modelling.

## S2.1. Exploration

To demonstrate a thorough exploration of at least parametric uncertainty, we point to Sepulveda et al. (2018). They assess the role of firm low-carbon generation in the future power system. They model 912 scenarios, combining several possible cases along seven aggregate dimensions of uncertain inputs. Though we stress that it is not necessarily the scale of their exploration that should be emulated, but their ethos. Their high and low cases in each dimension are extreme-but-plausible upper and lower bounds, rather than small divergences from a central estimate. Their approach is exhaustive, not perfunctory. Their exploration is thoughtful, building from a Global Sensitivity Analysis to determine the most influential parameters. Finally, they seek insights to inform decisions under all this uncertainty. For example, they conclude that, under any plausible outcome of the uncertainties they considered, firm low-carbon resources lower decarbonized electricity system costs. This is an insight for all uncertain futures, not a justification of insight for a central projection of the future.

As another example, we point the reader toward the work of Neumann and Brown (2023)[S2] who use multi-fidelity surrogate modelling to explore the dynamics of a widely used open-source European power system model, PyPSA-Eur. This surrogate approach allows them to explore over 50,000 optimisation scenarios, combining Global Sensitivity Analysis methods with Modelling to Generate Alternatives to observe the structure of the model's solution space across a range of uncertainties for technology capital costs. This approach explores aspects of both parametric and structural uncertainty. We further commend their transparent communication of study limitations.

These previous two examples are expansive computational explorations of parametric uncertainty. These types of works are more available in the literature, but we stress that the intention of the exploration principle is to extend considerations beyond modellable uncertainties. For example, Mersch, Markides, and Mac Dowell (2023)[S3], while exploring fuel price uncertainty (an in-model parametric exploration), acknowledge a wide range of uncertainty types in their "Limitations of the study" section. They raise parametric uncertainties within the model (such as technological learning rates), structural uncertainties of the model (a central decision-maker with perfect foresight), and unmodelled uncertainties (political instability and global energy markets), with some brief discussion of their potential implications for insights. These limitations are not obvious to readers until explicitly stated. While they may reduce confidence in the realism of the model, justification in Diagnostic Modelling does not depend on the realism of the model. These limitations bound the insights of the study, reducing misinterpretation and guiding further testing, which may lead to stronger justification (or falsification) in the future.

### S2.2. Interrogation

See our discussion of Cano Renteria, Schwartz, and Jenkins (2025) in the main text's *Interrogate models for testable real-world explanations* section.

### S2.3. Communication

Usher and Strachan (2012)[S4] perform a two-stage stochastic optimisation of the future UK power system. Their stochastic optimisation is limited to nine states of world due to computational tractability—a narrow exploration of uncertainty. However, they readily recognise these limitations and thoughtfully discuss their implications for decision-making throughout. They also focus discussion on real-world insights for decision making, justified through thoughtful mechanistic explanation rather than claims of model or study quality, and present these insights within their epistemic limits. For example, they highlight the value of short-lived demand technologies to quickly adapt to structural changes in energy supply, reducing the cost of uncertainty. A greater exploration of uncertainty would likely have yielded deeper or more robust insights, which they themselves admit, but the humility of their discussion exemplifies the Diagnostic Modelling philosophy and, notably, produced enduring insights that remain as relevant and valuable now as then, 13 years ago.

# S3. Uncertainty in energy systems modelling

Several methods have emerged to explore uncertainty in ESMs. These are each best suited to explore certain types of uncertainty so, to identify gaps, we explore a taxonomy of uncertainty for ESMs (Table S1) and map out the appropriate applications of existing uncertainty methods (Table S2).

Following established theory[S5,S6], we classify uncertainty first by ***aleatory*** (that which can be modelled but still not feasibly reduced) versus ***epistemic*** (that which is unknown or unknowable and therefore either cannot be modelled or can only be modelled via subjective Bayesian probabilities)—these being the ***resilience*** of the uncertainty—why it cannot be eliminated. To this category, we would add a third, more-mundane resilience: ***practical***—that which *could* feasibly be eliminated but not without trading off some limited resource (e.g., time, scope).

We further follow established practice for energy modelling[S7] in classifying uncertainty by ***parametric*** (input data) versus ***structural*** (design choices made in crafting the model), these being the ***insertion*** of the uncertainty—how it enters the model or study. To this category, we would again add a third, human insertion: ***subjective***—that which inserts at the transition between the computer model or reality and the mental model of the interpreter or model builder.

***Table S1. A proposed taxonomy of uncertainty for ESMs and some examples for each category of uncertainties common to ESM studies.*** *These categories are neither rigid nor exclusive; note how some uncertainties could span multiple categories. For example, a single parameter (say, the price of natural gas) can contain aleatory parametric uncertainty (stochastic variability), epistemic parametric uncertainty (undiscovered new gas reserves—or maybe there is simply no published price data), and epistemic subjective uncertainty (a failure to include it in uncertainty analysis). Aleatory uncertainty is often epistemic or practical in disguise, resulting from irreducible complexity. For example, wind speeds may be treated as aleatory if we lack either sufficiently detailed precision (epistemic uncertainty) or sufficiently powerful compute (practical uncertainty) to produce perfect forecasts of a complex climate system. We aim for clearly defined categories but must allow, then, for fuzzy categorisation.*

| | | **Insertion**<br>*How the uncertainty enters the model or study* | | |
|---|---|---|---|---|
| | | **Subjective**<br>Design/interpretation. Inserts at the transition from/to the mental model. | **Parametric**<br>Imperfect data. May be embedded in raw data or emerge in data processing. | **Structural**<br>Model structure. Embedded in structural design choices of the model. |
| **Resilience**<br>*Why the uncertainty cannot be eliminated* | **Practical**<br>We could eliminate this uncertainty but to do so would be costly. | ▪ Scenario design<br>▪ Choice of uncertainties to analyse<br>▪ Triage of model improvements<br>▪ Interpretation of results and communication of insights | ▪ Using available data instead of collecting or creating new data<br>▪ Internal consistency of modelled scenarios | ▪ Simplification, aggregation, scoping |
| | **Aleatory**<br>We can model this uncertainty but the uncertainty cannot be reduced. | ▪ Assumptions in place of complex systems (e.g., GDP growth rate)<br>▪ Assumptions of uncertainty distribution<br>▪ Conflicting stated preferences | ▪ Stochastic variables (e.g., wind speeds, streamflow, forced outages)<br>▪ Variability in future technology costs | ▪ Top-down representation (e.g., econometrics) |
| | **Epistemic**<br>We cannot rigorously model the uncertainty or we are not aware the uncertainty exists. | ▪ Failure to model all uncertainties<br>▪ Unconscious biases, assumptions, paradigms<br>▪ Inappropriate conflation of model results with reality<br>▪ Unrevealed preferences | ▪ Bad or missing data; data processing bugs<br>▪ Uncertainty in future mean technology costs<br>▪ Probability of future natural disasters | ▪ "Black box" model complexity<br>▪ Systems too complex to model faithfully<br>▪ Model code bugs |

Several uncertainty taxonomies already exist[S7], and we do not propose that ours should become the standard. Rather, we find that each taxonomy provides a valuable alternative framing and perspective to expand thinking about uncertainty. We have found this particular taxonomy useful to help identify less-quantifiable uncertainties, which are more difficult to model but no less important to recognise and discuss. As we map out this taxonomy, we see several uncertainties emerge for which a simple exploration of the parameter space may not be sufficient (e.g., deep parametric uncertainty, communication of insights) yet most ESM studies that perform any uncertainty analysis explore only parametric uncertainty[S7]. The rest of this matrix remains underexplored and underacknowledged.

***Table S2. Methods for uncertainty analysis in energy systems models.*** *Strengths and criticisms are not our conclusions but only claims we have found made or referenced in the literature. Please see our disclaimer at the beginning of Section S1.2.*

| Method | Brief description | Proposed strengths & applications | Criticisms | Examples |
|---|---|---|---|---|
| Global Sensitivity Analysis (GSA) [i] | Numerous statistical and probabilistic tools to determine the model input variables which most contribute to an interest quantity depending on model output[S8]. | Identify most influential inputs[S9]. N-at-a-time perturbation[S7]. Prioritise improvements and uncertainty analysis[S9]. Reduce model to influential variables (variance-based surrogate modelling) [S9]. Validating centroid results[S10]. | Can be computationally intensive[S10] (curse of dimensionality). May require subjective parameter bounds or uncertainties to explore[S7]. Implies complete exploration but not all uncertainties are modelled (i.e., typically only captures parametric uncertainties, or a subset thereof). | [S2,S11–S13] |
| Modelling to generate alternatives (MGA) | Any method used to systematically search the near optimal solution space for alternative solutions[S7]. | Finding near-optimal solutions that better satisfy unmodelled factors (balancing competing objectives)[S7,S14]. Reveal overlooked possible futures[S7]. Overcome bias of scenario construction[S7]. Unmask unrobust "knife-edge" solutions[S7]. Insights for structural uncertainties and correlations in the energy system. | Computationally intensive. Curse of dimensionality, the near-optimal space cannot be fully explored. Subjective slack value[S7]. Near-optimal space provides no measure of "goodness" for contained solutions[S7] (epistemically abstract). Subjectivity of exploration objective function. Does not "truly" explore structural uncertainty[S7]. | [S2,S14–S18] |
| Model inter-comparison | Comparing the outputs of models representing the same or similar systems, usually harmonising inputs to examine differences of outputs to reveal | Revealing subjective uncertainties like model structural choices, unconscious bias, and interpretation of results[S20]. Validating model results or insights[S19,S21,S22]. | Difficult and intensive to coordinate. Limited data points (number of models). Subjective choices and biases may be paradigmatic to the community (common | [S19,S20,S22–S25] |
| | model structural differences[S19]. | | to all models and therefore invisible). Harmonisation of inputs may mask true uncertainties. | |
| Robust decision-making (RDM) | Assess alternatives under myriad futures seeking satisfactory solutions that minimize regret[S26] to yield better decisions under conditions of deep uncertainty[S27]. | No need for input probabilities[S27]. Identification of robust alternatives under deep uncertainty[S27]. Exploratory and consensus-based[S27]. | Exploring many futures is intensive[S27]. Ambiguity of robustness-criterion[S27]. Sensitive to heuristic selection of "plausible" scenarios[S28]. | [S29,S30] |
| Robust optimisation (RO) | Optimising to minimise cost against adverse realizations of uncertain parameters within a given uncertainty set[S7]. | Computationally-tractable[S7]. No need for probability distributions[S7]. Model cost of hedging against uncertainty[S7]. Identify key hedging technologies[S7]. Quantify importance of uncertainty sources[S7]. | Bounds are subjective. Can only explore parametric uncertainties. Provides a single solution (limited information)[S7]. Decision approach is inherently minimax regret. | [S31–S35] |
| Scenario analysis | Using small ensembles of scenarios (narratives) or one-at-a-time (OAT) sensitivity analysis to explore alternative realisations of uncertainties[S7]. | Narratives are easy to communicate to policy-makers (high policy impact)[S7]. Low computational intensity. Easy to do. | Minimal exploration of uncertainty[S7]. Only first-order uncertainty, no compounding Nth order effects. Little to no treatment of tail risks[S7]. Misrepresents certainty[S7]. | See Yue et al. (2018)[S7] for examples. |
| Stochastic programming | Optimising considering multiple possible realisations of future uncertainties at once, typically with assigned probabilities to minimise expected cost[S7]. | Generate a hedged strategy to minimize expected cost for all possible futures[S7]. Identify hedging technologies or policies[S7]. Model cost of hedged planning and value of information[S7]. | Requires probabilities which can be themselves uncertain or subjective[S7]. Very computationally intensive, can only assess a few uncertainties[S7]. Result is a single hedged strategy, losing information of individual scenarios. | [S36–S42] |

[i] *Includes stochastic sampling methods like Monte Carlo, exploration methods like the Morris method, or Latin Hypercube Sampling (LHS), and variance-based methods like Sobol analysis.*

## S4. The curse of dimensionality

As a thought exercise, let us consider N independent uncertain variables. So, consider an N-dimensional unit hypercube, where each of these N dimensions represents the distribution of a single independent uncertain variable, taking one of three values with equal probability: {0 (low), 0.5 (central), 1 (high)}. This is a standard approach to representing uncertain variables in ESMs[S7,S27], where low, central, and high might be, for one common example, cheap, moderate, and expensive future natural gas prices, and “central” represents our best guess. Scenarios are then assembled from narratively cohesive combinations of these three cases for many uncertain variables, and a small ensemble of scenarios modelled to observe results under different “plausible” futures. Other, more-sophisticated uncertainty approaches like Stochastic Programming, Robust Optimisation, Monte Carlo, and Modelling to Generate Alternatives are gaining popularity, but scenario analysis remains the standard[S7].

Let us focus on an ensemble of all our central cases (which could be interpreted as our best guess outcome of all future uncertainty) and the expected divergence of all $3^N$ possible cases from this centre. In one dimension, a unit line with a centroid at 0.5, the farthest point from this central estimate is 0.5. In two dimensions, a unit square, we add two new points at a distance of 0.5 from the centroid but four (the corners) at a distance of 0.707. In three dimensions, a unit cube, we add two more at 0.5, eight more at 0.707, and eight new vertices at 0.866 (Figure S1).

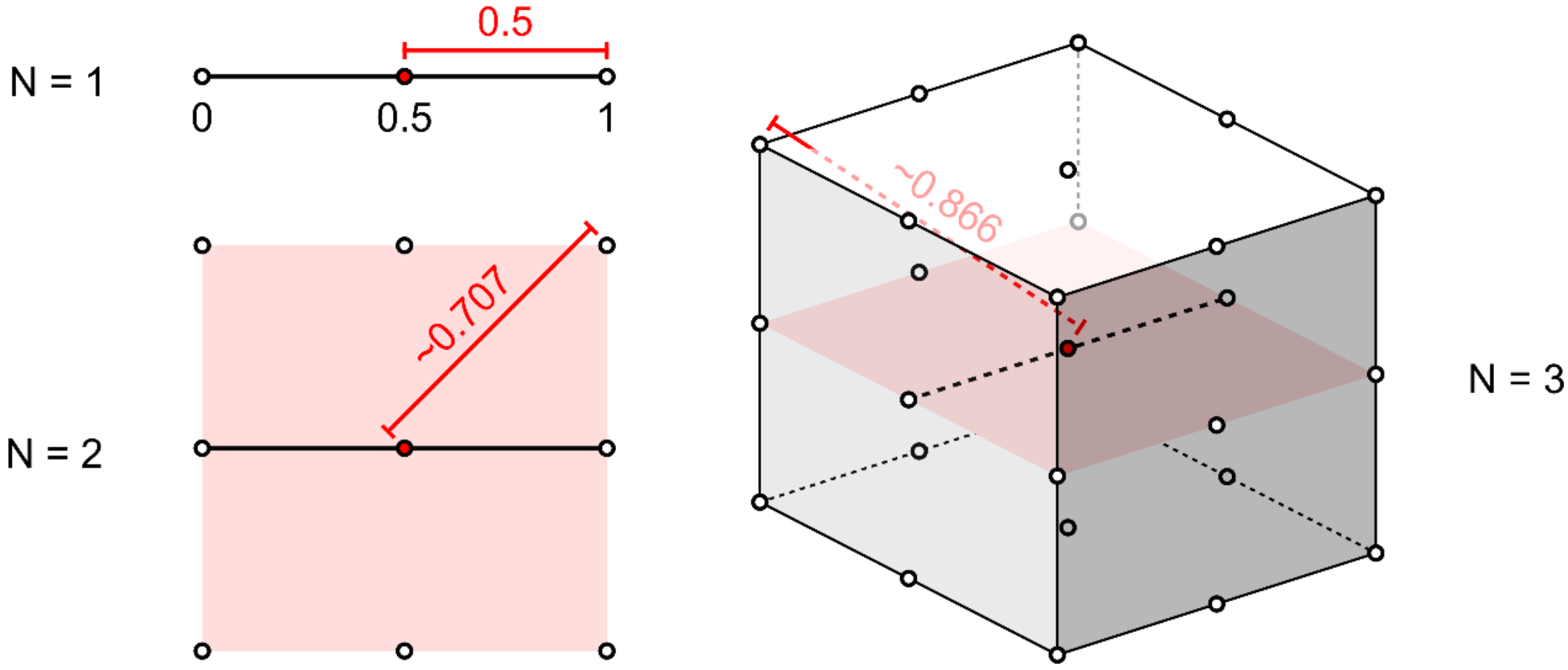

**Figure S1. Increasing uncertainty with increasing dimensionality.** As N increases (the number of uncertain variables or dimensions), additional corner points are introduced further from the centroid, increasing the mean displacement of possible realisations.

So, our furthest scenario from the centroid falls further away with each new variable/dimension. Further, with each new dimension, we add more points far from the centroid than near to it, drawing the expected divergence further away from our central scenario (Table S3).

**Table S3. New points added at various displacements from the centroid in increasing dimensions.** Each new dimension (each new uncertain variable) adds new possible scenarios further from the centroid, and adds more scenarios far from the centroid than near to it, increasing expected divergence from the centroid and therefore uncertainty. Note that, by 100 dimensions, there are $3^{100}$ or $10^{47}$ distinct points.

| N | Number of points at distance … from centroid | | | Farthest point | Mean distance from centroid |
|---|---|---|---|---|---|
| | 0.5 | 0.707 | 0.866 | | |
| 1 | 2 | - | - | 0.5 | 0.333 |
| 2 | 4 | 4 | - | 0.707 | 0.536 |
| 3 | 6 | 12 | 8 | 0.866 | 0.682 |
| … | … | … | … | … | … |
| 100 | 200 | 19,800 | 1,293,600 | 5 | 4.08 |
| 1000 | 2000 | 1,998,000 | 1,329,336,000 | 15.8 | 12.9 |

This is the paradoxical curse of dimensionality: the central region of the hypercube tends to "hollow out", with most points being found out at the extremities of the space in higher dimensions. This is also true for normal distributions or any other probability distribution we might reasonably sample from for each independent uncertain variable. The probability that all variables are randomly drawn near the centre decreases as more variables are drawn. The only escape is to limit the effective dimensionality of the uncertainty space, or find methods of justification that are compatible with this uncertainty.

# S5. Methods and tools for Diagnostic Modelling

We believe there is a need to develop and propagate new diagnostic tools for ESMs, and to better apply existing ones. These might be new modelling methods or protocols, new methods to explore uncertainty, or new methods of visualisation to communicate model dynamics. As a starting point, we have reviewed here a few available tools for Diagnostic Modelling and some examples of their application. This review is by no means exhaustive.

## S5.1. Tools to explore uncertainty

We review some common approaches to uncertainty exploration and examples of their application in Table S2. We note that a wealth of methods exists to explore aleatory parametric uncertainties while methods like Robust Decision-Making, Robust Optimisation, and Global Sensitivity Analysis are growing in popularity to explore epistemic parametric uncertainties.

Structural uncertainties are more challenging to explore as testing different structures often requires changing the mathematical formulation of the model. Though this is a topic of focus in the modelling community[S7,S43,S44], we cannot point to any particular tools for assessing general structural uncertainty, except perhaps model intercomparison. Modelling to Generate Alternatives can help to address one structural uncertainty, by revealing all the near-optimal solutions that might otherwise be hidden by single-solution optimisation. Lau et al. have made notable recent efforts to advance MGA methods to better explore this near-optimal space[S15,S18]. Similarly, Multi-Objective Optimisation has been utilised to explore trade-offs between competing alternative objectives, as demonstrated by Prina et al. (2020)[S45].

We see less explicit effort to develop rigorous methods to explore subjective uncertainties, though transparency, perhaps the most powerful tool against subjectivity, remains a central principle in the modelling community[S46–S48]. Modelling to Generate Alternatives helps to address one form of subjective uncertainty—competing or uncertain objectives—by offering near-optimal alternatives that might perform better on alternative objectives. Esser et al. (2025) demonstrate this application[S49].

Multi-Objective Optimisation is, again, a core tool for comparing competing objectives.

One approach to address both structural and subjective uncertainties is model intercomparison, whereby entire models can be compared, including all their embedded uncertainties. There are ongoing efforts in model intercomparison such as the long-running Stanford Energy Modelling Forum[S25,S50], and notable recent work by Schivley et al. (2025)[S24], though best practices for model intercomparison are not, as far as we have found, clearly established. Prominent past model intercomparisons have largely used input data alignment and comparable numerical outputs to isolate and focus on structural differences of models (structural uncertainty)[S21,S24,S50,S51]. While this approach may facilitate interpretation of those structural differences, it neglects the reality that models do not exist in a computational vacuum. Models and their studies are inextricable from the subjectivities of their builders, users, and audiences. The data alignment approach suppresses subjectivities around selection and processing of data. Comparison of numerical outputs (rather than insights) likewise suppresses subjectivities of study design and interpretation of results. Both of these subjectivities would be present in any independent study. Differences in numerical inputs may also be difficult to meaningfully interpret, compared to insights, as we have argued in this work. We see potential value for some future model intercomparisons to perform blind comparison of entire studies aligned only on questions, comparing final insights with no alignment of methods or data. Though, paradigms and limited data availability can lead to unintentional alignment of parameters, structures, and subjectivities that will remain unexplored regardless.

## S5.2. Tools to interrogate models

Tools for interrogation are those which help us to understand why the model behaves as it does, and whether this behaviour reflects real world systems or is just an artefact of abstraction. Sensitivity analysis remains a powerful tool, here. By exploring macroscopic trends of model response to changing inputs (sensitivity), interpretable insights can emerge. Mallapragada, Sepulveda, and Jenkins (2020)[S52] demonstrate this with their sensitivity analysis of breakeven cost (see their Figures 3, 6, and 7).

Another approach to interpreting the behaviour of a model is model reduction. The black box model is reduced to a simpler model (a surrogate model or emulator) that parsimoniously captures the behaviour of the black box. This reduced, parsimonious model can then be interpreted to understand the behaviour of the original[S53–S55]. This approach is particularly popular right now in the analysis of large language models[S56–S58]. Interpretable emulators have also been applied effectively in climate modelling[S59,S60] and extended to integrated assessment modelling[S61,S62]. Diagnostic emulation (emulation for the purpose of interpretability) strikes us as a valuable avenue for future research in ESMs. Surrogate models have been applied to ESMs but mostly, as far as we have seen, to reduce computational cost (especially in Global Sensitivity Analysis)[S11,S12]. Quick et al. (2024) provide one example where the structure of the surrogate model itself was interpreted to better understand the structure of the original model[S13]. Neumann and Brown (2023)[S2] demonstrate the use of surrogates for sensitivity analysis, using surrogates to model over 50,000 scenarios in a European power system model, mixing methods of Modelling to Generate Alternatives, Multi-Objective Optimisation, and Global Sensitivity Analysis to assess model dynamics. We particularly note the clear trade-offs of investment decisions demonstrated in their Figure 7.

The key bottleneck in the interrogation process is from model outputs into the mental model of the interpreter. That interpreter may be the original modeller or an audience of their work, whose role it is to challenge the proposed mechanistic explanation. In either case, visualisation plays an important role. A clear visualisation of model behaviour can lead to immediate intuition. Rather than overstep our expertise by attempting to define what a good visualisation should be, we instead enumerate a few examples of visualisations that we have found particularly intuitive for understanding model behaviour. The energy modelling field may benefit from a more comprehensive review of visualisation methods for model diagnosis.

Neumann and Brown (2023)[S2] present a visualisation for the feasibility of certain energy technologies within the near-optimal space, as the cost-slack of this space expands (see their graphical abstract and Figure 6). This demonstrates, at a glance, how technologies may become feasible alternatives as we depart from the cost-optimal solution, and how much flexibility there is in the solution space for selection of that technology. Their Figure 7 also clearly demonstrates trade-offs between certain technologies in the near-optimal space.

These same authors demonstrate in Neumann and Brown (2021)[S63] the use of correlation plots to visualise trade-offs of investment decisions between technologies (see Figure 6). Correlation plots are also used by Sinha et al (2024)[S17] to similar effect (see Figure 6). Of this second article, we also note their visualisation of near-optimal investment decisions over time in Figure 1.

Prina et al. (2020)[S45] perform a Multi-Objective Optimisation of a sector-coupled Italian power system model, oemof-moea. They produce Pareto fronts to visualise the trade-off of alternative objectives with total cost. Their Figure 6, visualising the trade-off between $CO_2$ emissions and total cost, is particularly intuitive.

Pickering, Lombari, and Pfenninger (2022)[S64] examine how certain system constraints affect the flexibility of other system design choices using SPORES, a variant of Modelling to Generate Alternatives where diversity is sought not only between technologies but also in space (i.e., not only what is built but also where). In their Figure 3, they visualise how the restriction of each outcome metric then restricts the range of possibilities for each other metric. Figure 3A is particularly intuitive for understanding the trade-offs of biofuel utilisation with demand electrification.

Finally, Ruggles et al. (2024)[S65] examine the impact of modelling a greater number of weather years on system adequacy. They find that modelling this increased weather variability leads to higher system costs, the cause of which is visualised in Figure 2. This figure intuitively shows this loss of adequacy and the types of decisions the model is making to attempt to reduce loss of load, and how these decisions differ between different scenarios of generation mix.

# S6. Decision-making methods and tools for the energy transition

## S6.1. A broad classification of methods and tools for decision-making

Decision-making methods can be broadly classified as: qualitative decision-making methods, quantitative decision-making methods, and computer models (top-down, bottom-up, and hybrid forms).

Qualitative decision-making methods use structured subjective assessment of available information to reach a decision, usually relying on sociotechnical theory, consensus, and expertise to justify that decision. Qualitative methods are often necessary when: uncertainty is unquantifiable, the decision cannot be reasonably or ethically represented with any numeric model, or there does not exist a clear consensus of objective, as is so often the case in planning the energy transition. Compared to quantitative methods, qualitative methods typically capture broader sociotechnical dimensions, though recommendations may be less concrete or actionable due to unreduced complexity, theoretical abstraction, and perceived lack of objectivity[S66,S67].

Quantitative decision-making methods seek, through rigorous numerical analysis, to render a number to a decision, leaving the final recommendation to a comparison of numbers between alternatives. This has some perceived benefits over qualitative decision making: (1) quantitative analysis is seen (perhaps falsely) as being more objective, repeatable, and credible[S46,S66,S68]; (2) quantitative methodologies may uncover more generalisable, mechanistic insights[S46,S69]; (3) it can be easier to make and execute a final decision when the burden of that decision has been offloaded to hard numbers[S66–S68]. Criticisms reflect these strengths: (1) any analysis, no matter how rigorous, is structured through the subjectivities and biases of the analysts, especially under uncertainty[S66,S70,S71]; (2) quantitative methods narrow the scope to quantifiable dimensions, often excluding or minimising important but less-quantifiable dimensions of a decision[S66,S67,S72,S73]; (3) decision makers may overestimate the authority of quantitative recommendations, taking for granted the certainty, quality, and relevance of underlying analysis and data[S46,S66–S68,S70,S71,S74].

Computer models, a subset of quantitative methods where numerical analysis can be done effectively only by computer, feature many of the same benefits and criticisms of quantitative decision-making methods, amplified. Computer models are typically used (not to insist they *should* be used) when the dimensions of the decision, and therefore its representative numeric model, are too complex to be tractable by human analysts, but not so complex as to be out of reach of the current state of computer technology. The use of computers allows for the quantitative analysis of far more-complex decisions but, especially as computer technology has continued to advance (potentially exacerbated by growing ML/LLM integration), the meaning of that analysis—and its strengths and weaknesses—becomes increasingly opaque to the overseeing analysts (the "black box" problem)[S46,S67,S75,S76].

The top-down versus bottom-up paradigm of computer models can be explained as a difference in strategy between attempting to directly fit a model to observations of macroscopic system behaviours (top-down) versus attempting to model the underlying mechanisms of those behaviours (bottom-up)[S77–S79]. This parallels the paradigms of qualitative versus quantitative, where top-down models typically capture a broader range of system behaviours—intrinsic to macroscopic trends[S77–S79]—that might be lost in a more mechanistic model, while bottom-up models capture more-granular details of a narrower scope of interactions, retaining greater ability to analyse the mechanisms themselves[S77,S80]. Top-down studies are typically more "pessimistic", implicitly accounting for the hidden costs and barriers of the energy transition and reporting higher emissions abatement costs. Bottom-up studies are typically more "optimistic", allowing for a greater range of abatement options and assuming greater efficiency than is realistic, resulting in lower or even negative abatement costs[S77,S79–S81]. Hybrid models attempt to pair these paradigms, usually coupling top-down economic or behavioural models with bottom-up technological, agent-based, or environmental models[S77,S79,S80]. As computer models are limited by available resources like processing power, processing time, memory, data storage, and analysts' time and effort, analysts must carefully assess study priorities in crafting the scope of the model[S47,S66].

## S6.2. A review of methods and tools for decision-making

Our approach for this review was simply to research methods and tools that we had encountered in proximity to our own work, and then to explore additional methods and tools we encountered while reviewing literature on those. We searched for information to fill these tables by three main methods: web search using Google Scholar (terms like "<method> limitations", "<method> criticisms", "<method> strengths", or "<method> review"), following citations up and downstream from these articles, and conversation with an LLM (OpenAI's o3-deep-research via ChatGPT), asking for peer-reviewed literature that discussed these methods and tools, and following up for more authoritative or relevant articles. LLMs were used only to discover literature sources, which were then read by authors. All content in these tables was written directly by the authors after reading cited articles; no content was generated by LLMs.

We acknowledge and emphasise that this was not a rigorous or unbiased review process. The objective of this review was not to capture all available decision-making tools, nor was it to faithfully represent those reviewed here. Rather, these tables summarise a personal effort to expand our own view of the decision-making space and to better understand the position of our work in the broader scientific consensus. Strengths and criticisms are not our conclusions but only claims we have found made or referenced in the literature. We present criticisms to better understand perspectives seen in literature, but these do not necessarily reflect our own perspectives. We disagree with many of these criticisms and expect that other readers will, as well. In many cases, the cited articles are themselves summarising perceived criticisms that their authors may not agree with. This information should be viewed critically. We challenge readers not to accept this information as we have presented it here but rather to use this as an example to make a similar effort for themselves.

***Table S4. Review of some qualitative (or mixed-methods) decision-making methods for the energy transition.*** *Excluded from this review are the many sociotechnical frameworks[S82] that, in practice, typically inform and structure the application of these methods. Strengths and criticisms are not our conclusions but only claims we have found made or referenced in the literature. Please see our disclaimer at the beginning of Section S1.2.*

| Method | Brief description | Proposed strengths & applications | Criticisms | Examples |
| --- | --- | --- | --- | --- |
| Case study | Intensive, holistic analysis of a single or collective case[S83–S85] to understand a broader class of similar cases[S84]. | Flexibility[S83]. Depth of analysis[S84]. Descriptive (retrospective) insights[S84]. | Perceived low credibility[S83,S84]. Ambiguous methodology[S83–S85]. May lack generalisability[S86]. | [S87–S90] |
| Content analysis | A systematic, replicable technique for compressing many words of text into fewer content categories based on explicit rules of coding[S91]. | Finding trends and patterns in documents[S91]. Systematic analysis of text[S92]. | Interprets text too granularly (misses latent structures)[S93]. Sensitive to biases of analyst[S92] and poor methodology[S93]. | [S94–S96] |
| Expert elicitation | Formal procedures for obtaining and combining expert judgments[S97], typically to quantify uncertainties[S97,S98]. | Feasible under any degree of uncertainty or data availability[S97,S98]. Strong treatment of complexity[S98]. | Too subjective (overconfidence, bias of experts)[S98]. Challenging to validate[S97]. Resource-intensive process[S97]. | [S99–S101] |
| Process tracing | An analytical method for establishing causal relationships | Compatibility with high-n quantitative study. Retrospective | Confusion with storytelling. Difficulty in establishing causal | [S104,S105] |

| | between events in time based on qualitative case descriptions[S102]. | analysis to identify causes and factors influencing socio-technical transitions or policy development.[S103] | inference. Uncertainty in starting point.[S103] | |
|---|---|---|---|---|
| Qualitative comparative analysis (QCA) | For a set of cases, identify parsimonious sets (configurations) of casual variables as necessary or sufficient to achieve outcomes, assessed qualitatively using Boolean logic[S106]. | Bridge between low-n case-study and high-n quantitative study[S107–S109]. Complex inference of parsimonious causal relationships with low-N[S107,S108]. Evaluate competing explanations[S108]. Replicability[S109]. Equifinality[S107,S110]. | Sensitive to assumptions[S108]. May risk P-hacking[S108]. Weak causal inference[S111,S112]. Lacking rigour of traditional statistics[S111,S112]. | S109,S110,S113–S116 |
| Semi-structured interview | Qualitative data collection strategy in which the researcher asks informants a series of predetermined but open-ended questions[S117]. | Focussed but explorative[S118,S119]. Understanding of unique perspectives[S118]. Answering complex social-behavioural questions[S118]. | Credibility depends on sample[S118]. Analysis is subjective[S119,S120]. Interviews are unreliable[S120]. Limited validation, replicability[S120]. | S82,S121–S125 |
| Stakeholder engagement | A strategic process of interacting with stakeholders to gather information about shared interests, preferences, and the potential for joint action[S126]. | Promotes justice, participation, accountability, trust, and legitimacy[S127]. Considers preferences[S127]. Challenges academic thinking[S128]. Improves uptake and impact[S128]. Mitigates risk of public opposition[S129]. | When done poorly, alienates participants, erodes trust[S127]. Often performative, non-inclusive, ineffectual[S127], opaque[S130]. Conflicts arise between stakeholders[S128,S131]. Stakeholders' stated preferences may not match revealed preference[S131]. | S94,S129,S132–S134 |

***Table S5. Review of some quantitative decision-making methods for the energy transition.*** *This is by no means an exhaustive list and there is significant overlap between these methods and with qualitative methods. Strengths and criticisms are not our conclusions but only claims we have found made or referenced in the literature. Please see our disclaimer at the beginning of Section S1.2.*

| Method | Brief description | Proposed strengths & applications | Criticisms | Examples |
|---|---|---|---|---|
| Breakeven analysis [i] | Analysis of the point at which costs and benefits are equal[S135]. | Help develop targets and incentives[S136]. Compare alternatives[S137]. | Difficult to quantify compounding uncertainties[S135,S138]. May mistreat non- | S136,S140,S141 |
| | | Feasibility assessment[S137]. Overcome uncertainty[S135]. Identify data gaps[S138]. | monetary factors[S135,S138,S139]. | |
| Cost-benefit analysis (CBA) [i] | Comparing total costs to total benefits in monetary terms[S139]. | Quantify net benefits of alternatives[S139]. Directly compare diverse alternatives with different units of non-monetary costs or benefits[S139]. Enables a single common metric across multiple impact categories. | Difficult to parametrise[S138,S142,S143]. Dangerous to assign values under deep uncertainty (overconfidence)[S142]. May mistreat or undervalue non-monetary factors[S135,S138,S139]. | S143–S146 |
| Cost-effectiveness analysis (CEA) [i] | The evaluation of alternatives according to the ratio of their incremental costs to their incremental effects with regard to efficiently achieving some outcome[S147,S148]. | Choose most cost-efficient alternative to achieve an outcome, maximising outcomes for given cost[S147,S149]. | Ambiguity over discounting and which costs to include[S149]. Difficult to assess effectiveness[S149,S150] and uncertainties[S151]. May neglect co-benefits and externalities[S151]. | S152–S154 |
| Multi-criteria decision-making (MCDM) [ii] | Making decisions in the presence of multiple competing objectives[S155]. | Handling multiple, competing objectives[S156,S157]. Transparency of option selection[S156]. Practicality for choosing from alternatives[S156]. Solving complex problems[S156]. | Subjective weighting of criteria[S158]. Limited validity and interpretability of aggregate scores[S158]. | S157,S159–S161 |
| Multi-objective optimisation (MOO) [ii] | Searching for the best solution within a set of possible solutions when there exist multiple, irreducible competing objectives[S162]. | Explore and visualise decision trade-offs between competing objectives[S163,S164]. Allow for preferences in final decision, improving acceptability[S163]. Identify pareto-optimal alternatives[S163]. | Difficult to quantify non-monetary objectives[S164]. Can be computationally-expensive[S162]. Only provides efficient alternatives, not a final decision[S162]. Ambiguity of dominance under uncertainty[S165]. | S45,S165–S170 |
| Trade-off analysis (TOA) [iii] | A holistic framework to quantify and understand the | Assessing diverse and competing objectives[S171]. Integrating multi- | Subjective, often imbalanced selection and treatment of non-economic | S173–S176 |
| | trade-offs between competing objectives, typically between economic benefits and environmental harms, with spatial explicitness[S171]. | disciplinary research (participation)[S171]. Spatially-explicit trade-offs[S171]. Policy relevance[S172]. | objectives[S172]. Conflict between numeric models and qualitative methods[S172]. Reliant on availability of input data[S172]. | |

[i] There is significant overlap between these methods. Some works group these together as "cost analysis" methods along with cost-utility analysis and cost-feasibility analysis[S147].

[ii] MOO and MCDM inherently operate in tandem. MCDM requires MOO (explicitly or implicitly) to identify non-dominated alternatives and MOO requires MCDM (explicitly or implicitly) to evaluate and select from those alternatives[S163,S177].

[iii] TOA is a decision-making framework that typically applies several quantitative and qualitative methods in tandem.

***Table S6. Review of some top-down computer model types for energy transition decision making.*** *Notably absent are advanced artificial intelligence (AI) models[S178], the pieces of which are still settling into place—epistemologically speaking[S179]—after their explosion into widespread use in late 2022. Strengths and criticisms are not our conclusions but only claims we have found made or referenced in the literature. Please see our disclaimer at the beginning of Section S1.2.*

| Method | Brief description | Proposed strengths & applications | Criticisms | Examples |
|---|---|---|---|---|
| Computable general equilibrium (CGE) | Model the whole economy in circular closed-form, assuming equilibrium throughout, built on foundations in microeconomics. Optimisation maximises consumer utility and firm profits[S180,S181]. | Short-run shock simulation[S181]. Analysis of fiscal policy (e.g. carbon tax) and quantity instruments (e.g. emissions limit)[S180–S182]. Welfare analysis[S180,S181]. Price changes[S180,S181]. Inter-industry substitution[S181]. | Dependent on assumptions and calibration[S180,S181,S183]. Dependent on often out-dated parameters and IO/SAM data[S180,S181,S184]. Not appropriate for long-run analysis[S181,S182,S184]. Ill-suited to assess monetary policy[S181]. Poor treatment of social and environmental factors[S185], technological change, and the financial sector[S181]. Overly simplified[S183,S185]. | [S186–S189] |
| Econometric relationships [i] | Statistical inference (regression) to develop a model of economic phenomena based primarily on empirical | Long-run forecasting of historical economic trends[S191]. Descriptive analysis of policies' historical impacts on economic outcomes[S192,S193]. | Ill-suited to analyse structural change (e.g., policies) [S191,S194,S195]. Sometimes poorly grounded in economic theory[S191,S194]. | [S192,S193,S196–S198] |

| | historical data[S190]. | | Overly-aggregated agents[S194]. Data dependent[S180]. Can be a black box[S180]. Correlational relationships may not be causal. | |
|---|---|---|---|---|
| Input-output (I-O) | A linear matrix model tracing the flow of goods and services[S199] within the economy of a given region, from initial input through intermediate use to end use, aggregated by economic sector[S200]. | Captures interdependence of all economic sectors[S201]. Linear (tractable)[S201,S202]. Multi-regional flows analysis[S203]. Often adapted to account for sectoral resource consumption and environmental impacts[S201–S204]. | Does not capture substitutional effects[S201, ii]. Dependent on static and potentially out-dated I-O tables[S201]. Restricted by aggregation of available I-O data[S202]. | S203,S205–S207 |
| Regression | Statistical inference of relationships between variables to build a predictive model of the effect of one variable upon another[S208]. | Describing relationships[S209]. Extrapolating or interpolating relationships[S209]. Treatment of aleatory uncertainty[S210]. Identifying causal mechanisms[S211–S213]. | Relies heavily on subjective assumptions[S214,S215]. Sensitive to data quality, extreme values[S214]. Can be used to make false inference or imply causality[S214,S215]. | S96,S133,S212,S216,S217 |
| System dynamics (SD) | Representing a system as a network of interdependent stocks, flows, and variables with feedback loops and delays over continuous time[S218]. | Handling non-linearities, complex dynamics, path dependence[S219,S220]. Including time delays[S219,S220]. Representation of actors[S219]. Representation of socio-political factors[S221]. Flexibility for mixed-methods[S220]. | Models may be too abstract[S221]. Forecasts are often inaccurate[S222]. Structure and parameters depend on subjective assumptions[S223]. Impossible to objectively validate[S224]. | S220,S225–S227 |

[i] Modern macroeconometric models (DSGE, VAR) have strived to define structures using established economic theory, such as rational expectations[S228] or IS-LM[S181], to improve representation of economic shocks and therefore ability to forecast responses to policy change. However, these models have faced similar criticism as older regression models in their poor representation of agent behaviour[S194,S195,S229] and poor performance in predicting economic outcomes[S194]. Romer (2016) suggests that the included microfoundations are essentially performative and primarily serve to cloak the shortcomings of models behind black box complexity, lending them undue credibility[S229].

[ii] Much like CGE models, substitutional effects can be incorporated into I-O models via own- and cross-price elasticities (dynamic input-output models[S201]). Further dynamics can be incorporated, such as supply-side allocation functions[S201], but down this road the line between I-O modelling and CGE modelling is quickly blurred.

***Table S7. Review of some bottom-up computer model types for supporting energy transition decision making.*** *Strengths and criticisms are not our conclusions but only claims we have found made or referenced in the literature. Please see our disclaimer at the beginning of Section S1.2.*

| Method | Brief description | Proposed strengths & applications | Criticisms | Examples |
|---|---|---|---|---|
| Agent-based (ABM) | Simulates the dynamics of complex systems based on describing the continuous interaction of agents who are heterogeneous in terms of information and decision rules[S230]. | Accounts for agent heterogeneity (preferences, bounded rationality, difference of information, social interactions)[S230,S231]. Assessing non-market factors (e.g., information) and their interactions with market factors[S230]. | Complex (compounding uncertainties, computationally intractable)[S230]. Difficult to parametrise, calibrate, and validate[S230]. | [S230,S232–S234] |
| Earth system (ESM) [i] | Global climate models with the added capability to explicitly represent biogeochemical processes[S235]. | Forecast greenhouse gas concentrations[S235]. Assess climate engineering options[S236]. Understanding human impact on the environment[S236]. Identifying feedback loops[S235,S236]. | Extremely computationally intensive; significant simplifications often needed[S235]. Difficult to validate[S235]. Poor data availability[S235]. Conflicts of spatial scale[S235]. High uncertainty[S230,S237,S238]. | [S237,S239,S240] |
| Energy systems optimisation (ESOM) [ii] | Detailed, bottom-up technology models, typically utilising linear programming techniques to minimize the system-wide cost of energy provision (or maximise surplus) by optimizing the installation of energy technology capacity and its utilisation[S47]. | Long-term strategic planning (decarbonisation pathways)[S46,S47,S241,S242]. Flexible spatial, temporal, and technological resolution[S241,S242]. Evaluating technologies and policies[S241,S242]. Technological fidelity[S46,S241]. | Complex and difficult to interpret, lacking transparency[S43,S46–S48,S241–S244]. Computationally intensive, requiring simplification[S46,S241,S243]. Large, uncertain input datasets[S43,S47,S241,S243]. Optimistic decision representation (optimisation, rationality, perfect information, efficient markets)[S77,S241,S243]. Cost as primary objective misrepresents real- | [S241,S246–S250] |

| | | | world decision-making[S74,S245]. | |
|---|---|---|---|---|
| Geographic information system (GIS) | A system for capturing, storing, checking, manipulating, analysing, and displaying data which are spatially reference to the Earth[S251]. | Linking together and overlaying diverse geospatial datasets[S251,S252]. Geospatially-explicit inputs to other models[S253]. Siting and technoeconomic analysis (renewables[S252,S253], transmission infrastructure[S252]). Land-use planning[S254,S255]. Environmental modelling[S256,S257]. Visualisation. | Poor data availability, lack of standard data formats.[S253,S257] Difficult to harmonise datasets at different spatial resolutions[S253,S257]. Often stuck with inaccurate geospatial data[S257]. | S252,S258–S260 |
| Life cycle assessment [iii] (LCA) | The compilation and evaluation of the inputs, outputs, and the potential environmental impacts of a product system throughout its life cycle[S261]. | Holistic, comparative impact assessment (energy, emissions and other environmental harms, and resource consumption)[S262,S263]. Trade-offs across impact categories[S262]. Analysing process hot-spots for targeted policies[S262]. | Inconsistent data and methodology[S262–S264]. Lack of transparency in methodology[S262,S264]. Disconnect between practitioners and decision makers, lack of participation[S262,S264]. Data intensive[S263,S264]. Large uncertainties[S263,S264]. Subject to manipulation and arbitrary rules (e.g., allocation). | S265–S267 |
| Techno-economic assessment (TEA) | A method to evaluate economic performance of technological systems based on the underlying design of the system[S268]. | Technology screening, and selection[S269–S271]. Process synthesis and optimisation[S272]. Network design[S273,S274]. Projection of future technology costs[S275]. | Inconsistent methodology. Lack of transparency in methodology[S276,S277]. Cost metrics of limited relevance to real-world or system level decision making[S278,S279]. | S280–S284 |

[i] A broader classification might be environmental modelling. This umbrella term includes atmospheric models but also hydrological, air pollution, ecosystem/land-use models and others. It would not be reasonable to review these diverse model types as one category and they are too numerous to review individually within the scope of this analysis, so, we have included only the all-encompassing earth system model.

[ii] For practical reasons, we have excluded the broad category of simulation models, being those which attempt to simulate the evolution of an energy system over time. This would include everything from electricity market models to urban energy system models to simulation models of plant-level operations like industrial facilities, solar or wind farms, as well as demand models, scenario-based projections, or even stock turnover models. The types and applications are myriad and diverse, as are the decisions they inform.
[iii] Including IO-LCA, where an input-output matrix is used to account economy-wide for sectoral environmental impacts (especially emissions).

***Table S8. Review of hybrid computer model types for energy transition decision making.*** *The distinction between EEMs and IAMs is not so clear in practice[S285]. Here, we make the distinction at the inclusion of environmental modelling, though this could again be considered a middle ground (energy-environment-economy, E3) where IAMs require the further inclusion of social factors. Strengths and criticisms are not our conclusions but only claims we have found made or referenced in the literature. Please see our disclaimer at the beginning of Section S1.2.*

| Method | Brief description | Proposed strengths & applications | Criticisms | Examples |
|---|---|---|---|---|
| Energy-economy (EEM) | An energy planning model which deploys a combination of bottom-up technological explicitness and top-down micro/macro-economic behaviours[S286]. | Assess technology-focussed policy with behavioural realism[S287]. Combine detailed technological representation and long-run energy system changes with realistic costs and macro-economic feedbacks[S288,S289]. | Inconsistency of behavioural assumptions between paradigms[S286,S289,S290]. Conflicting granularity[S290]. Complexity (black box, computationally intensive)[S286]. Data intensive[S286]. Often poor handling of uncertainty[S287]. | [S287–S289,S291–S294,S294] |
| Integrated assessment (IAM) | Computer models that describe the potential long-term evolution of the global energy system, as well as other GHG-emitting systems such as agriculture and land use[S295], integrating elements from various disciplines such as engineering, economics, climate, and land use[S296]. | Cross-disciplinary insights[S297–S300]. Long-term projections at a global scale[S298–S300]. Simulate mitigation pathways and policies[S298–S300]. Holistically assess social and economic harms of climate change and quantify social cost of carbon[S299,S300]. | Impossible to validate[S74,S295,S297]. Lack of available data[S74,S295,S301]. Oversimplification of complex systems[S74,S295,S296,S298,S301,S302]. Too much uncertainty to explore[S74,S295,S301,S302]. Poor treatment of tail uncertainties[S74,S295,S301,S302]. Reliance on expert elicitation for probabilities[S74,S301]. Disconnect with policy makers[S74,S295,S297,S302]. Complexity (black box, computationally intensive)[S295,S297,S301]. Unrealistic, cost-optimal | [S21,S303–S307] |

| | | | paradigm[S74,S245,S295,S296,S298,S302]. | |
|---|---|---|---|---|

# Supplemental References